\documentclass[sn-mathphys-num]{sn-jnl}

\usepackage{graphicx}%
\usepackage{multirow}%
\usepackage{amsmath,amssymb,amsfonts}%
\usepackage{amsthm}%
\usepackage{mathrsfs}%
\usepackage[title]{appendix}%
\usepackage{xcolor}%
\usepackage{textcomp}%
\usepackage{manyfoot}%
\usepackage{booktabs}%
\usepackage{algorithm}%
\usepackage{algorithmicx}%
\usepackage{algpseudocode}%
\usepackage{listings}%
\usepackage{physics}
\usepackage{hyperref}
\usepackage[capitalise]{cleveref}

\begin{document}

\title[Article Title]{Temperature-driven transition between momentum-resolved and disordered averaged Coulomb drag in 1D systems}


\author[1]{\fnm{Mingyang} \sur{Zheng}}

\author[1]{\fnm{Rebika} \sur{Makaju}}

\author[1,2]{\fnm{Rasul} \sur{Gazizulin}}
\author[3]{\fnm{Alex} \sur{Levchenko}}
\author[4]{\fnm{S.J.} \sur{Addamane}}
\author*[1]{\fnm{Dominique} \sur{Laroche}}\email{dlaroc10@ufl.edu}

\affil*[1]{\orgdiv{Department of Physics}, \orgname{University of Florida}, \orgaddress{\city{Gainesville}, \postcode{32611}, \state{Florida}, \country{USA}}}

\affil[2]{\orgdiv{National High Magnetic Field Laboratory High B/T Facility}, \orgname{University of Florida}, \orgaddress{\city{Gainesville}, \postcode{32611}, \state{Florida}, \country{USA}}}

\affil[3]{\orgdiv{Department of Physics}, \orgname{University of Wisconsin–Madison}, \orgaddress{\city{Madison}, \postcode{3706}, \state{Wisconsin}, \country{USA}}}

\affil[4]{\orgdiv{Center for Integrated Nanotechnologies}, \orgname{Sandia National Laboratories}, \orgaddress{\city{Albuquerque}, \postcode{87185}, \state{New Mexico}, \country{USA}}}





\maketitle

\section*{Abstract}

Advancing the understanding of electron–electron interactions in one-dimensional systems remains one of the central challenges in low-dimensional physics, especially for Coulomb-coupled Tomonaga-Luttinger liquids. Notably, the difficulty of reliably extracting one-dimensional system parameters, combined with the presence of disorder, has hindered the interpretation of 1D Coulomb drag experiments. Here, we present a self-consistent experimental determination of the relative Luttinger liquid interaction parameters through 1D Coulomb drag measurements, and achieve quantitative agreement with theoretical predictions. Utilizing vertically coupled GaAs/AlGaAs quantum wires, we fully characterize the one-dimensional parameters through magnetic depopulation. Coulomb drag exhibits a systematic evolution with magnetic field, reflecting the successive depopulation of 1D subbands and the suppression of disorder effects. Two distinct temperature regimes are identified, marking the boundary between momentum-resolved and disordered-averaged Coulomb drag. The observed scaling, peak broadening, and nonlinear current–voltage characteristics establish a unified and quantitative framework for probing electron–electron interactions in 1D systems. 

\section*{Main}\label{sec1}

Electron–electron interactions lie at the heart of condensed-matter physics, and their consequences are most striking in low-dimensional systems. Even the simple repulsive Coulomb interaction can give rise to a wealth of emergent collective phenomena, including correlated insulating states and superconductivity \cite{chen_evidence_2019, guo_crossover_2020, li_wigner_2024}. Despite the rich physics of low-dimensional systems, quantitatively measuring electron–electron interactions remains challenging, particularly in one dimension where the conventional quasiparticle picture breaks down, and interacting electrons are described by the Tomonaga–Luttinger liquid (TLL) framework, in which low-energy excitations are collective bosonic modes of spin and charge \cite{Tomonaga_1950, Luttinger_1964, Haldane_1981}. Tunneling spectroscopy and tunneling conductance measurements can probe the band structure of collective excitations and reveal power-law behavior within isolated 1D systems; however, they do not directly probe inter-system interactions \cite{bockrath_luttinger-liquid_1999, auslaender_spin-charge_2005, jin_momentum-dependent_2019, weldeyesus2025dominant}. Coulomb drag, in contrast, provides a powerful and direct probe of these interactions \cite{narozhny_coulomb_2016}. In a drag experiment, a current driven through one conducting system induces a measurable voltage or current in a nearby, electrically isolated system via Coulomb coupling alone \cite{narozhny_coulomb_2016}. Beyond its broad use in detecting excitons in bilayers \cite{Kellogg_2003, eisenstein_boseeinstein_2004, liu_quantum_2017, qi2025perfect, nguyen2025perfect}, Coulomb drag has emerged as a uniquely sensitive tool for studying interactions in 1D systems \cite{flensberg_coulomb_1998, klesse_coulomb_2000, debray_experimental_2001, pustilnik_coulomb_2003, fuchs_coulomb_2005, peguiron_temperature_2007, Levchenko_2008, dmitriev_coulomb_2012, laroche_positive_2011, laroche_1d-1d_2014, makaju_nonreciprocal_2024, zheng2025tunable, zheng2025quasi, zheng2025spin}.

Thus far, quantitative comparisons between theory and experiments have remained elusive \cite{klesse_coulomb_2000, fuchs_coulomb_2005, laroche_1d-1d_2014, zheng2025tunable}, owing to technical challenges in device fabrication and measurements, as the drag signal is exponentially suppressed with increasing interwire separation $d$ \cite{fuchs_coulomb_2005, narozhny_coulomb_2016, zheng2025tunable}. In addition, recent experiments have revealed a nonreciprocal Coulomb drag signal, which violates Onsager reciprocity and lies beyond the conventional momentum-transfer picture \cite{makaju_nonreciprocal_2024, fu2025non, zheng2025tunable}. Although such nonreciprocal drag is qualitatively consistent with charge fluctuation theories, its observation underscores that 1D Coulomb drag remains incompletely understood and calls for further theoretical and experimental investigation \cite{Levchenko_2008, narozhny_coulomb_2016}.  

In this work, we employ vertically coupled quantum wires fabricated from bilayer GaAs/AlGaAs quantum wells with an interwire separation of only 15 nm to systematically investigate interactions between density-mismatched TLLs in a perpendicular magnetic field. By measuring the temperature- and magnetic-field-dependent Coulomb drag, we observe a momentum-transfer-dominated drag signal that oscillates as a function of magnetic field. The drag signal exhibits a power-law temperature dependence at intermediate temperatures (Mid-T) and an exponential temperature dependence at high-temperatures (High-T). By measuring the magnetic depopulation of the quantum wires, we quantitatively characterize the electron density and effective width of electrostatically defined quantum wires for the first time in a Coulomb drag experiment, enabling a direct and precise comparison between the experimental work and theoretical predictions. Our observations are fully consistent with the suppression of Coulomb drag with increasing Fermi wave vector ($k_F$) mismatch and with the thermal broadening of the drag response at elevated temperatures \cite{fuchs_coulomb_2005}, and enable the identification of the transition between momentum-resolved and disordered averaged Coulomb drag.

\subsection*{Device operation and characterization}\label{sec2}

The device used in this work is fabricated from a molecular beam epitaxy--grown GaAs/AlGaAs electron bilayer heterostructure consisting of two 18-nm-wide GaAs quantum wells separated by a 15-nm-wide Al$_{0.3}$Ga$_{0.7}$As barrier. A representative optical image and a schematic of the vertically coupled quantum wire device are shown in Figs.~\ref{fig1}a and \ref{fig1}b, respectively. The top and bottom wires are formed in the upper and lower two-dimensional electron gas (2DEG) layers, giving a vertical interwire separation of $d_\mathrm{vert}=33~\mathrm{nm}$. The two wires are independently contacted and electrostatically tunable, as described in the Methods. 

For Coulomb drag measurements, a current $I_\mathrm{Drive}$ is driven through the bottom wire (drive wire), while both the longitudinal resistance of the drive wire $R^{\mathrm{Drive}}_{xx}$ and the drag voltage $V_\mathrm{Drag}$ induced in the top wire (drag wire) are measured under a perpendicular magnetic field $B$. Reversing the direction of the drive current leads to a sign reversal of $V_\mathrm{Drag}$, consistent with the Onsager reciprocity relations (Fig.~\ref{fig1}) \cite{Onsager_1931}. In contrast, exchanging the roles of the drive and drag wires breaks reciprocity (Supplementary Fig.~S6 and Fig.~S7). As a tunable crossover between reciprocal and nonreciprocal drag has been demonstrated previously \cite{zheng2025tunable}, the present work focuses on the configuration in which disorder creates a nearly symmetric potential that preserves current-reversal reciprocity, leading to the observation of a momentum-transfer induced drag signal and preserved current-reversal reciprocity. The nonreciprocal case is briefly discussed in Supplementary note 2.

Figures~\ref{fig1}c and \ref{fig1}d show two-dimensional maps of the drag voltage as a function of the drive- and drag-wire gate voltages; the insets display line cuts taken at $V_\mathrm{BPL}=-0.3~\mathrm{V}$ (line~1), explicitly showing the reciprocity of the drag signal. For momentum-transfer–dominated Coulomb drag, electrons in the drive and drag wires move in the same direction, producing a negative drag voltage, as shown in Fig.~\ref{fig1}c. A pronounced checkerboard pattern is observed in both drag maps, arising from the enhancement of the drag signal near the opening or closing of 1D subbands in either wire. This enhancement results from the combined effects of the diverging 1D density of states, reduced Fermi velocity, weakened screening, and efficient interwire backscattering near subband edges, which maximize momentum transfer \cite{klesse_coulomb_2000, debray_experimental_2001, pustilnik_coulomb_2003, laroche_positive_2011, laroche_1d-1d_2014, narozhny_coulomb_2016}. Similar subband-edge enhancement is also expected for nonreciprocal drag, where the strongest asymmetry occurs near the opening of a new conductance channel \cite{Levchenko_2008, zheng2025tunable}. We note that the drag signal reaches a maximum along the boundary corresponding to the first subband of the drag wire.

Figures~\ref{fig1}e and \ref{fig1}f show the drag resistance $R_\mathrm{Drag}=-V_\mathrm{Drag}/I_\mathrm{Drive}$ as a function of the top plunger gate voltage $V_\mathrm{TPL}$ along line~1 at $0~\mathrm{T}$ and $7.2~\mathrm{T}$, respectively. At zero magnetic field, the drag resistance is maximized near $V_\mathrm{TPL}=-1.35~\mathrm{V}$, where the first subbands of the drive and drag wires align. Weaker peak features associated with higher subband openings are also observed at larger gate voltages. At $7.2~\mathrm{T}$, the perpendicular magnetic field depopulates the higher subbands, leaving only the first subband occupied in both wires. Small modulations superimposed on the main drag peak are attributed to disorder. Taken together, these signatures unambiguously identify the observed signal as momentum-transfer–dominated Coulomb drag between quantum wires, and rule out contributions from capacitive coupling or Coulomb drag mediated by quantum dots \cite{Keller_2016, sierra2019fluctuation}.

\subsection*{Magnetic depopulation}\label{sec3}

For an electrostatically defined 1D channel under a perpendicular magnetic field, additional magnetic confinement is added to the gate-defined electrostatic potential, producing hybrid magnetoelectric subbands (see Methods). Similar to the quantum Hall effect, the perpendicular magnetic field produces three key effects in 1D wires: increased subband spacing, enhanced effective mass with reduced kinetic energy, and successive magnetic depopulation of the 1D subbands \cite{berggren_magnetic_1986,van_houten_magnetic_1987}. As successive hybrid subbands depopulate, the wire conductance exhibits magnetic-field induced oscillations with conductance minima corresponding to the Fermi energy crossing subband edges as shown in Fig.~\ref{fig2}b for both the bottom and top wires at $V_{\mathrm{TPL}}=0~\mathrm{V}$. The characteristic confinement frequency $\omega_0$, wire width $W$, and 1D carrier density can be extracted by fitting the magnetic depopulation oscillations using an electro-magnetic parabolic confinement model \cite{Berggren_1986, berggren_magnetic_1986, berggren_magnetic_1986-1, berggren1988characterization}. Figure~\ref{fig2}c shows representative oscillations extracted from Supplementary Fig.~S10 at $V_{\mathrm{TPL}}=-0.5~\mathrm{V}$. 

Magnetic depopulation was measured for both wires along line~1 over a wide range of top plunger gate voltages $V_{\mathrm{TPL}}$ from $0$ to $9.1~\mathrm{T}$, beyond which the wires are fully depleted. Figure~\ref{fig2}a shows a two-dimensional map of the top (drag) wire conductance as a function of $V_{\mathrm{TPL}}$ and magnetic field, where clear subband depopulation is visible. The transition from two-dimensional to one-dimensional transport occurs at $V_{\mathrm{TPL}}\approx -0.38~\mathrm{V}$. By systematically fitting the magnetic depopulation of both wires along line~1, we extract the top wire width and the 1D carrier densities $N_e^{1D}$ of the top and bottom wires as functions of $V_{\mathrm{TPL}}$, as shown in Figs.~\ref{fig2}g and \ref{fig2}h. As $V_{\mathrm{TPL}}$ is decreased from $-0.5~\mathrm{V}$ to $-1.4~\mathrm{V}$, the top wire is gradually pinched off, with its width shrinking from approximately $220~\mathrm{nm}$ to $120~\mathrm{nm}$ and its 1D density decreasing from $4\times10^8~\mathrm{m^{-1}}$ to $1\times10^8~\mathrm{m^{-1}}$. Over the same gate range, the bottom wire density decreases more weakly, from $5\times10^8~\mathrm{m^{-1}}$ to $\sim3.5\times10^8~\mathrm{m^{-1}}$, due to interwire electrostatic crosstalk. Comparing the top wire subband structure in Fig.~\ref{fig1}e with the extracted density in Fig.~\ref{fig2}h, we identify the 1D density of the first subband of the top wire to be approximately $1.6\times10^8~\mathrm{m^{-1}}$ and around four subbands in total at $0~\mathrm{T}$.

The Coulomb drag resistance measured along line~1 is shown as a function of magnetic field in Fig.~\ref{fig2}e. With increasing magnetic field, the first subband drag peak shifts toward more positive gate voltages and exhibits an overall increase in amplitude, accompanied by pronounced oscillations. At higher fields, the drag resistance is gradually suppressed and eventually vanishes for $B\gtrsim9~\mathrm{T}$, where the drag wire becomes fully depleted, as summarized in Fig.~\ref{fig2}f. Comparison with the magnetoconductance of the drag and drive wires measured at fixed gate voltages (Fig.~\ref{fig2}b) shows that minima in the drag oscillations coincide with conductance minima of the drag wire. Notably, the drag maxima do not occur at these depopulation points, but instead appear on either side, coinciding with conductance maxima. Although the density of states diverges at a subband edge, the longitudinal Fermi velocity simultaneously vanishes, leading to inefficient momentum transfer and a minimum in drag resistance. When the Fermi level lies slightly above the subband bottom, the density of states remains enhanced while carriers retain finite kinetic energy, resulting in maximal interwire momentum transfer and pronounced drag peaks. As subbands are successively depopulated with increasing magnetic field, reduced screening enhances the effective interwire Coulomb interaction, accounting for the overall growth of the drag signal. The alternating peak amplitudes reflect the spin structure of the depopulation process, with larger peaks arising from spin-degenerate Landau-level-like subbands and smaller peaks appearing once Zeeman splitting resolves the individual spin branches. The final drag peak near $7~\mathrm{T}$ is weaker than expected from this trend, as the carrier density becomes too small for the enhanced density of states to compensate the reduced number of mobile electrons, leading to a suppression of the drag signal as the wire approaches pinch-off. The magneto-oscillations are consistent with a generalized saddle potential model described in the Methods. Although this model is more suitable for short, gated wire constrictions-such as quantum point contacts \cite{Levchenko_2008}, we find that it nevertheless captures the essential qualitative physics of field-induced oscillations in the multi-mode regime characteristic of our long wires. The results of this analysis are plotted in Fig. \ref{figGdrag}.

\subsection*{Temperature dependence}\label{sec4}

To identify the dominant scattering mechanism of Coulomb drag in a single subband at high magnetic fields, we analyze the temperature dependence of the drag signal in Fig.~\ref{fig3}. Fig.~\ref{fig3}a shows the result for magnetic fields ranging between 0 and $7.2~\mathrm{T}$ in the single subband limit. For all magnetic fields, the drag resistance increases as the temperature approaches zero, a characteristic signature of interacting TLLs \cite{klesse_coulomb_2000, pustilnik_coulomb_2003, fuchs_coulomb_2005}. For magnetic fields $B \geq 1.1~\mathrm{T}$, a characteristic transition temperature $T_1$ separates the temperature dependence into an intermediate-temperature regime and a high-temperature regime. In the Mid-T (High-T) regime, the drag resistance decreases (increases) with increasing temperature. In contrast to previous studies without a magnetic field, where $T_1$ was reported to be around $1.5~\mathrm{K}$ \cite{laroche_1d-1d_2014, zheng2025tunable, zheng2025spin}, we find that $T_1$ exceeds our measurement limit of $3.2~\mathrm{K}$ at $0~\mathrm{T}$ and decreases approximately linearly with increasing magnetic field, as shown in Fig.~\ref{fig3}b. The temperature dependence of $R_\mathrm{Drag}$ is analyzed using both log–log and Arrhenius representations, as shown in Figs.~\ref{fig3}c and \ref{fig3}d. Although both regimes can be fitted in either representation over limited ranges, the Mid-T regime exhibits a substantially larger linear range and a higher coefficient of determination $R^2$ in log–log scale (Supplementary Table~S1), whereas the High-T regime shows superior linearity in the Arrhenius plot.

The independently characterized density mismatch between the two wires and the observed transition from power-law to Arrhenius behavior naturally motivate comparison with the theory of Coulomb drag between quantum wires with unequal electron densities \cite{fuchs_coulomb_2005}. In the spin-polarized regime, the linear drag resistivity is given by
\begin{equation}
\rho_D = \rho_0\left(\frac{T}{E_0}\right)^{4K-3} f(Q/T,K),
\end{equation}
where $E_0$ is of order the Fermi energy, $K\equiv K_{c}^{-}$ is the relative TLL charge interaction parameter ($0<K<1$ for repulsive interactions), and $Q \equiv \hbar v_F \delta k_F / k_{b} K$ quantifies the Fermi-momentum mismatch between the two wires. The nonlinear drag regime can be excluded here due to the low drive current of $2~\mathrm{nA}$, and current–voltage characteristics are discussed in Fig.~\ref{fig4} and the methods. For $Q=0$, the drag resistivity follows $\rho_D \propto T^{4K-3}$. For $\lvert Q\rvert \gg T$, the drag signal is exponentially suppressed, and the drag resistivity takes the form $\rho_D \propto T^{-1} e^{-Q/T}$ so that $\rho_D(Q)$ peaks at $Q =0$ with a width proportional to the temperature \cite{fuchs_coulomb_2005}. We note that, in the spin-full regime, the power-law exponent is expected to scale to $2K-1$ \cite{klesse_coulomb_2000}.

A representative result from the fitting procedure along line~1 at $3.3~\mathrm{T}$ is shown in Figs.~\ref{fig3}e and \ref{fig3}f across the first-subband drag peak. At this field, the drag resistance is maximized near $V_\mathrm{TPL}\approx -1.3~\mathrm{V}$, where the first-subband depletion points of the drive and drag wires align (Fig.~\ref{fig3}g). Results at different fields are presented in Supplementary Figs.~S12 and S13. As the magnetic field increases beyond $\sim 1.7~\text{T}$, the system transitions from the spinfull regime to the spin polarized regime. In the Mid-T spinfull regime, the majority of the extracted power-law exponents yield unphysical null or negative $K$ values. Similar results were previously reported both in laterally-coupled \cite{makaju_nonreciprocal_2024, cai2026nonreciprocalcoulombdragballistic} and vertically-coupled devices \cite{zheng2025tunable, zheng2025spin} in a regime where reciprocal and nonreciprocal drag coexist. These results point towards the ineffectiveness of the standard theoretical formalism for 1D Coulomb in the presence of disorder. 

In the spin-polarized regime, power-law exponents compatible with conventional theories for 1D Coulomb drag are obtained. For density-mismatched wires, the drag peak occurs at minimal momentum mismatch $Q$, corresponding here to $V_\mathrm{TPL}\approx-1.3~\mathrm{V}$ at $3.3~\mathrm{T}$. In the Mid-T regime, the drag resistance near this peak follows the power-law scaling $\rho_D\propto T^{4K-3}$ expected for $Q\approx0$, yielding a TLL interaction parameter $K\approx0.5$ from the observed slope of approximately $-1$. As the gate voltage moves away from the drag peak and $Q$ increases, the Mid-T power-law exponent exhibits an upturn, reflecting a reduced decrease rate of the drag resistance with temperature and an effective increase of the extracted interaction parameter toward $K\approx0.75$. In contrast with the increase of reduced screening, the larger $K\approx0.75$ near pinchoff is not readily explained by theoretical models and likely arises from the interplay between disorder and density mismatch.

In the High-T regime, the temperature dependence is governed by exponential suppression arising from density mismatch, $\rho_D\propto T^{-1}e^{-Q/T}$. For an Arrhenius exponent of approximately $-2$ and $K=0.5$, theory predicts a 1D density mismatch $\delta n_{1D}=1.93\times10^5~\mathrm{m^{-1}}$ for a drag-wire single-subband density $n_{1D}=1.6\times10^8~\mathrm{m^{-1}}$. The condition $Q/T\gg1$ required for this exponential suppression is satisfied predominantly in the High-T regime near $T\sim2~\mathrm{K}$. Consistent with increasing momentum mismatch away from the drag peak, the Arrhenius exponent shows a corresponding downturn. Phenomenologically, the drag resistance exhibits a slower decrease in the Mid-T regime and a more rapid increase in the High-T regime on both sides of the drag peak as temperature rises. This behavior is consistent with the experimentally observed increase of the drag peak width with temperature (Fig.~\ref{fig3}h), which is a key theoretical prediction for density-mismatched wires \cite{fuchs_coulomb_2005}. 

The emergence of a crossover temperature $T_1$ at relatively high temperatures ($\sim 1~\mathrm{K}$) is not predicted by standard momentum transfer theories, however. Indeed, the predicted increase of drag resistance associated with the formation of interlocked Wigner crystals is only expected at temperatures $\lessapprox 10~\mathrm{mK}$ \cite{klesse_coulomb_2000}. Instead, we attribute this crossover to disorder effects, which are largely neglected in most theories of 1D Coulomb drag \cite{klesse_coulomb_2000, fuchs_coulomb_2005}. While the Arrhenius dependence $e^{-Q/T}$ originates from momentum-mismatch suppression, the crossover to a power-law behavior can be related to enhanced disorder-induced backscattering at $T<T_1$, which relaxes momentum conservation \cite{fiete_coulomb_2006} and suppresses the effectiveness of momentum mismatch. As temperature increases, disorder effects diminish and become negligible for $T>T_1$, where the Arrhenius dependence becomes dominant. $T_1$, which is proportional to the disorder potential, is around $3~\mathrm{K}$ at $0~\mathrm{T}$ in our device. The enhanced disorder effect at low temperature has been widely observed in 1D systems, including the suppression of conductance quantization in disordered quantum wires \cite{Tarucha_1995, PhysRevB.50.11008} and temperature-tunable reciprocal and nonreciprocal Coulomb drag \cite{zheng2025tunable}. The approximately linear decrease of $T_1$ with magnetic field further supports this interpretation: as $B$ increases, stronger electro-magnetic confinement narrows the wire and suppresses scattering from charged impurities and edge roughness, thereby reducing the effective disorder strength. This magnetic-field induced disorder reduction is also consistent with the transition from unphysical interaction parameters extracted at low fields to the reasonable interaction parameter values extracted at larger fields.

\section*{Conclusions}\label{sec6}

We identify a momentum-transfer–dominated Coulomb drag regime in vertically coupled, non-ballistic quantum wires, evidenced by pronounced drag peaks at the alignment of drag- and drive-wire subbands. Using gate control, we systematically tune and characterize the one-dimensional wire width and carrier density, enabling direct comparison with theoretical descriptions of 1D Coulomb drag in density-mismatched systems. A perpendicular magnetic field strongly enhances the drag signal and induces oscillations arising from magnetic depopulation of 1D subbands. The high data quality allows us to resolve distinct temperature regimes, with power-law and Arrhenius dependences governing the intermediate- and high-temperature behavior, respectively. Our results elucidate the physical origin of the crossover temperature $T_1$ as a transition from a disorder-dominated momentum relaxation regime to a momentum-resolved regime where drag is suppressed by Fermi-momentum mismatch. In the momentum resolved regime, the observed density-dependent exponents, temperature-dependent peak broadening, and nonlinear current–voltage characteristics are in excellent agreement with the theory of Coulomb drag between quantum wires with unequal electron densities \cite{fuchs_coulomb_2005} (see Methods for more details on nonlinear drag measurements). These findings position vertically coupled quantum wires as a highly controllable platform for investigating interaction-driven transport in one dimension. More broadly, the demonstrated control of Coulomb drag through proximity, confinement, and magnetic field provides a foundation for engineering interaction-mediated functionalities in future low-dimensional nanocircuits, such as novel topological phases \cite{Schrade_2017}, high-throughput entangled electrons \cite{Ueda_2018}, and heat harvesting \cite{roche_harvesting_2015}.

\backmatter

\section*{Methods}\label{sec7} 

\bmhead{Material growth and device fabrication}
The wires were patterned on an n-doped GaAs/AlGaAs electron bilayer heterostructure with two 18-nm-wide quantum wells separated by a 15-nm-wide Al$_{0.3}$Ga$_{0.7}$As barrier. The unpatterned combined density and mobility of the GaAs quantum wells are $n = 2.98 \times 10^{11} \, \text{cm}^{-2}$ and $\mu = 7.4 \times 10^4 \, \text{cm}^2 / \text{V} \cdot \text{s}$, respectively. After a mesa-structure was wet-etched using phosphoric acid, Ge–Au–Ni–Au ohmic contacts were deposited on the structure and annealed at 420 $^{\circ}$ for 60 s. A set of two Ti–Au split gates was then defined on the surface of the heterostructure, using electron-beam lithography [Fig. \ref{fig1}a]. Once the upper side processing was completed, bare GaAs was epoxied on top of the heterostructure and the sample flipped, mechanically lapped and chemically etched using subsequent citric and hydrofluoric mixtures until the lower 2DEG was only $\sim$ 150 nm away from the lower surface (now on top of the device), following the EBASE technique \cite{weckwerth_epoxy_1996}. To ensure that no off-mesa leakage occurred between the bottom and top gates, a thin 40 nm layer of Al$_2$O$_3$ was deposited on the new surface using atomic layer deposition. Using phosphoric acid, vias were then etched through the surface to enable electrical connection to the ohmic contacts and the split gates buried under the surface of the device. Finally, using electron-beam lithography, another set of Ti–Au split gates was defined on the top side of the device, and aligned with the bottom gates. The end result is depicted in Fig. \ref{fig1}a, with additional optical images provided in Supplementary Note 1.

\bmhead{Wire confinement and tunability}

Each quantum wire is electrostatically defined by a pair of pinch-off (PO) and plunger (PL) gates. The top and bottom wires are independently contacted through ohmic contacts to the upper and lower layers, respectively, by biasing the corresponding PO gates. The PL gate primarily controls the electrostatic confinement of each wire, tuning both the effective wire width and the 1D carrier density \cite{laroche_positive_2011, laroche_1d-1d_2014}. Because the two wires are vertically separated by only $d_\mathrm{vert}=33~\mathrm{nm}$, the plunger gates also produce finite interwire electrostatic crosstalk. This crosstalk shifts the depletion boundaries in two-dimensional gate maps and is taken into account when comparing drag features with the conductance of the individual wires. 

\bmhead{Measurement techniques}

Transport measurements were performed in a dilution refrigerator (LD250, Bluefors) with a base temperature less than 7 mK. The device was mounted in an experimental cell, which is thermally anchored to the mixing chamber of the dilution refrigerator. The polycarbonate cell is then filled with liquid Helium-3. The liquid is thermalized to the mixing chamber of a dilution refrigerator via annealed silver rods that enter the Helium-3 cell. These bring the system’s base temperature near that of the dilution unit, below 15 mK. All measurements were performed in an ultraquiet environment, shielded from electromagnetic noise. RC filters with cutoff frequencies of 50 kHz were employed to reduce RF heating. All measurements were performed using standard low-frequency lock-in ampliﬁcation techniques. Additionally, source-measure units were used to source and measure DC signals applied to the electrostatic gates. Transport measurements on individual quantum wires were performed at base temperature using a constant 100 $\mu$V excitation at 13 Hz in both wires in a two-contact configuration. The Coulomb drag measurements, unless otherwise specified, were performed in a constant-current mode where a 2 nA current was sent at 13 Hz through the drive wire. In this configuration, the out-of-phase current was always much smaller than the in-phase current. The drag voltage is measured across the drag wire while both sides of the drag wire are virtually grounded, as shown in Supplementary Fig.~S2. Two-dimensional drag maps are acquired by sweeping the plunger gate voltages of the drive and drag wires and recording $V_\mathrm{Drag}$ at each gate-voltage pair.

In these maps, the drag signal reaches a maximum along the boundary corresponding to the first subband of the drag wire and vanishes upon complete depletion of the drag wire. The curved boundary in the two-dimensional drag maps arises because each gate sweep is terminated when the drive wire pinches off, with the curvature reflecting interwire electrostatic crosstalk. Corresponding two-dimensional conductance maps of the drive and drag wires are shown in Supplementary Fig.~S5 and align with the drag-signal boundaries.

The tunneling measurements, as shown in Supplementary Note 2, were performed by sending a small source-drain voltage across the device for different top PO gate and bottom PO gate values. The combination was selected such that the tunneling resistance between the two wires was larger than 10 M$\Omega$ in a bias range of $\pm 3 \text{mV}$. Detailed configuration of the drag measurement is also presented in the Supplementary Note 2.

\bmhead{1D wires in perpendicular magnetic field}

For an electrostatically defined one-dimensional (1D) channel in a GaAs quantum-well two-dimensional electron gas (2DEG), the lateral confinement induced by negatively biased gates can be approximated by a parabolic potential $V_0(x)=m^*\omega_0^2x^2/2$ and result in 1D subbands with energies $E_n=\hbar\omega_0(n+1/2)$, where $\omega_0$ is a characteristic confinement frequency and $m^*$ is the electron effective mass \cite{berggren_magnetic_1986}. Applying a perpendicular magnetic field $B$ introduces both an additional magnetic parabolic confinement $V_B(x)=m^*\omega_c^2(x-x_0)^2/2$ and Landau-level quantization. Here, $\omega_c = eB/m^{\star}$ is the cyclotron frequency and $x_0=\hbar k/eB$ is the guiding-center coordinate with $k$ as the 1D wavevector and $e$ as the electron's charge. The combined electrical and magnetic confinement yields a stronger effective parabolic potential with characteristic frequency $\omega=(\omega_0^2+\omega_c^2)^{1/2}$ and hybrid electro-magnetic subbands with dispersion
\begin{equation}
E_n(k)=\hbar\omega(n+1/2)+\frac{\hbar^2k^2}{2m^*(B)},
\end{equation}
where the magnetic-field-dependent effective mass is $m^*(B)=m^*\omega^2/\omega_0^2>m^*$.

Under the perpendicular magnetic field, the conductance plateaus of the top wire oscillate and evolve into well-quantized plateaus near $3.3~\mathrm{T}$, as shown in Fig.~\ref{fig2}d. Similar quantization is observed in the bottom wire and is not attributed to mixing with the Hall resistance $R_{xy}$, as verified by reversing the magnetic field (Supplementary Fig.~S8). In fact, such quantization is observed only for the gate voltage corresponding to the 2D regime at 0~T, where 1D wires are not yet fully formed. It results from the edge channel reflection due to the different quantum Hall filling numbers in the 2D reservoirs and 1D wires. A similar effect has been observed in the $R_{xx}$ across the quantum point contact (QPC) in a magnetic field \cite{van1988four}. 

\bmhead{Theoretical model and simulation}

In this model, transport through the quantum drag circuit in each wire is described in terms of energy-dependent transmission eigenvalues for each mode within the Landauer picture. Unlike conductance, which is determined by the transmission eigenvalues themselves, drag is determined by their differences at slightly shifted energies. This reflects the sensitivity of drag to the breaking of particle-hole symmetry, which is most pronounced near conductance plateau transitions. Therefore, sweeping the gate voltage across the transition between two consecutive plateaus leads to a spike in drag conductance.

Incorporating a magnetic field into this model alters the confining and gate-induced scattering potentials for electrons, ultimately modifying the transmission coefficients. At least qualitatively, these effects can be described using transmission coefficients derived previously in Refs. \cite{Fertig_1987,Buttiker_1990} for a saddle-point constriction. By combining these results with the expressions for drag conductance from Ref. \cite{Levchenko_2008}, we extracted its field dependence. The results of this analysis are plotted in Fig. \ref{figGdrag}. To generate these plots, we fixed the gate voltage in one wire and varied it in the other to enable a multi-mode regime under the magnetic field. As the varied gate voltage opens more channels and the magnetic field is swept, the drag conductance exhibits several pronounced oscillatory features, similar to the experimental data. However, that simple model fails to capture the non-monotonic dependence of the oscillation magnitude on the magnetic field. A more precise model including explicit electron-electron interactions would likely be needed to reproduce this effect. We note that, for a weak drag signal, the drag resistivity and conductivity are linearly related: $\rho_{Drag} = \frac{\sigma_{Drag}}{\sigma_{1}\sigma_{2}}$. Here, $\sigma_{1}, \sigma_{2}$ are the conductivities of the drive and drag wire, respectively.

\bmhead{Nonlinear drag measurements and analysis}

Nonlinear drag measurements were performed by sweeping the drive current in the range $1~\mathrm{nA}\leq I_\mathrm{Drive}\leq30~\mathrm{nA}$ along line~1 for magnetic fields between $0~\mathrm{T}$ and $7.2~\mathrm{T}$. Following our previous work \cite{zheng2025tunable, zheng2025quasi}, we extract the symmetric and antisymmetric components of the drag voltage by reversing the drive current direction, defined as $V^\mathrm{S}_\mathrm{Drag}=(V^\mathrm{R}_\mathrm{Drag}+V^\mathrm{L}_\mathrm{Drag})/2$ and $V^\mathrm{AS}_\mathrm{Drag}=(V^\mathrm{R}_\mathrm{Drag}-V^\mathrm{L}_\mathrm{Drag})/2$, respectively. These correspond to the reciprocal (momentum-transfer) and nonreciprocal drag contributions. The dependence of $V^\mathrm{AS}_\mathrm{Drag}$ and $V^\mathrm{S}_\mathrm{Drag}$ on $I_\mathrm{Drive}$ along line~1 at $0~\mathrm{T}$ is shown in Figs.~\ref{fig4}a and \ref{fig4}b, respectively. Across all subband fillings and drive currents, $V^\mathrm{AS}_\mathrm{Drag}$ is approximately five times larger than $V^\mathrm{S}_\mathrm{Drag}$, confirming that the drag signal in our device is dominated by momentum transfer. This dominance is further supported by the near-zero symmetric–antisymmetric ratio $r_\mathrm{SAS}$ observed at both $0~\mathrm{T}$ and $7.2~\mathrm{T}$ (Supplementary Fig.~S16). Over the entire current range, the positions of the drag peaks show negligible dependence on $I_\mathrm{Drive}$, ruling out peak-shift-induced nonlinearities. Figures~\ref{fig4}c and \ref{fig4}d show representative nonlinear $I$–$V$ curves extracted at four $V_\mathrm{TPL}$ values indicated by arrows in Figs.~\ref{fig4}a and \ref{fig4}b, corresponding to drag-wire subband occupancies $N_\mathrm{Drag}=0.5$, $1$, $2$, and $>2$, from left to right. For $N_\mathrm{Drag}=2$ and $N_\mathrm{Drag}>2$, both $V^\mathrm{AS}_\mathrm{Drag}$ and $V^\mathrm{S}_\mathrm{Drag}$ are predominantly linear, with only weak nonlinearity appearing in $V^\mathrm{AS}_\mathrm{Drag}$ at $I_\mathrm{Drive}>20~\mathrm{nA}$. As $V_\mathrm{TPL}$ approaches the first-subband drag peak, the nonlinearity becomes increasingly pronounced and evolves into a nonmonotonic drag voltage at $V_\mathrm{TPL}=-1.3~\mathrm{V}$.

This nonlinear and nonmonotonic behavior is consistent with the theory of nonlinear Coulomb drag in density-mismatched quantum wires developed by Fuchs, Klesse, and Stern \cite{fuchs_coulomb_2005}, which predicts nonlinear and nonmonotonic $I$–$V$ characteristics for Tomonaga–Luttinger liquid interaction parameters $K=0.75$ and $K=0.25$, respectively. In particular, for $K=0.25$ and $K=0.5$, the theory predicts that the drag voltage decreases with increasing drive current once $\sqrt{2}\hbar I_\mathrm{Drive}/(ek_BQ)\sim1.14~\text{and}~1.2$, corresponding to a transition current $I^0_\mathrm{Drive}$ of $34~\mathrm{nA}$ and $36~\mathrm{nA}$ for a momentum mismatch $Q=2$, respectively. A smaller momentum mismatch $Q$ leads to a lower $I^0_\mathrm{Drive}$ and a larger drag voltage near the drag peak. Consistent with this expectation, we observe that for $V^\mathrm{AS}_\mathrm{Drag}$ at $N_\mathrm{Drag}=1$, the drag voltage begins to decrease with increasing drive current at $I^0_\mathrm{Drive}\approx20~\mathrm{nA}$ (Fig.~\ref{fig4}b). Moreover, $I^0_\mathrm{Drive}$ increases as $V_\mathrm{TPL}$ moves away from the drag peak and the momentum mismatch $Q$ increases, in agreement with the theoretical prediction \cite{fuchs_coulomb_2005}. We note that the elevated current used in the nonlinear regime likely smears out the disorder effect and enables our wire to be described by a momentum-transfer model for clean wires.

Despite the nonlinear $I$-$V$ characteristics, the drag resistance $R_\mathrm{Drag}=-V_\mathrm{Drag}/I_\mathrm{Drive}$ exhibits an approximately linear dependence on $I_\mathrm{Drive}$ in the high-bias regime ($I^0_\mathrm{Drive}>10~\mathrm{nA}$), as shown in Figs.~\ref{fig4}c and \ref{fig4}d. In the low-bias regime, $R_\mathrm{Drag}$ remains independent of $I_\mathrm{Drive}$, consistent with linear-response behavior. In the high-bias regime, $R_\mathrm{Drag}$ is well described by a linear form,
\begin{equation}
R_\mathrm{Drag}=R^0_\mathrm{Drag}+k\,I_\mathrm{Drive},
\end{equation}
with representative fits shown in Figs.~\ref{fig4}g and \ref{fig4}h. Comparing the fitted slope $k$ and intercept $R^0_\mathrm{Drag}$ with the low-bias drag resistance measured at $I_\mathrm{Drive}=2~\mathrm{nA}$ (Fig.~\ref{fig1}e), we find that both quantities reproduce the fine structure associated with the first-subband depletion of the drag wire and disorder features in the drive wire. This correspondence implies that the high-bias slope is proportional to the low-bias drag resistance, $k=c\,R^0_\mathrm{Drag}$. Consequently, the drag resistance in the high-bias regime can be written as
\begin{equation}
R_\mathrm{Drag}=R^0_\mathrm{Drag}\left(1+c\,I_\mathrm{Drive}\right),
\end{equation}
with a proportionality constant $c\approx-0.02~\mathrm{nA}^{-1}$ that is independent of $V_\mathrm{TPL}$. As shown in Figs.~S14 and S15, these results are also observed in the spin-polarized regime.



\section*{Data availability}\label{sec8} 

The data that support the findings of this study are available within the paper and its Supplementary Information. The raw data that support the findings will be available in Zenodo upon publication.

\section*{Code availability}\label{sec9} 

The code used for analysis will be available in Zenodo upon publication.

\bibliography{citations}

\section*{Acknowledgments}\label{sec9} 


We acknowledge the helpful discussions with Z. Lu. This work was supported by the National Science Foundation through NSF/DMR-2518016. This work was performed, in part, at the Center for Integrated Nanotechnologies, an Office of Science User Facility operated for the U.S. Department of Energy (DOE) Office of Science. Sandia National Laboratories is a multimission laboratory managed and operated by National Technology $\And$ Engineering Solutions of Sandia, LLC, a wholly owned subsidiary of Honeywell International, Inc., for the U.S. DOE’s National Nuclear Security Administration under contract DE-NA-0003525. The views expressed in the article do not necessarily represent the views of the U.S. DOE or the United States Government. Part of this work was conducted at the Research Service Centers of the Herbert Wertheim College of Engineering at the University of Florida.  A portion of this work was also performed at the National High Magnetic Field Laboratory and was partially supported by the National High Magnetic Field Laboratory through the NHMFL User Collaboration Grants Program (UCGP). The National High Magnetic Field Laboratory is supported by the National Science Foundation through NSF/DMR-1644779 and NSF/DMR-2128556 and the State of Florida. A. L. acknowledgments support by the National Science Foundation Grant No. DMR-2452658 and H. I. Romnes Faculty Fellowship provided by the University of Wisconsin-Madison Office of the Vice Chancellor for Research and Graduate Education with funding from the Wisconsin Alumni Research Foundation.

\section*{Author contribution}\label{sec10} 
M.Z., R.M., and D.L. fabricated the sample. M.Z. and R.G. performed the measurements of the vertically coupled devices. S.J.A. performed the growth of the double quantum well heterostructures. D.L. designed and supervised the experiments. M.Z. and D.L. analyzed the data. A.L. developed the theoretical model. M.Z., A.L., and D.L. co-wrote the Letter, and all authors discussed the results and commented on the manuscript.

\section*{Competing interests}\label{sec11}

The authors declare no competing interests.


\begingroup
\begin{figure}[h!]
\centering
\includegraphics[width=1\textwidth]{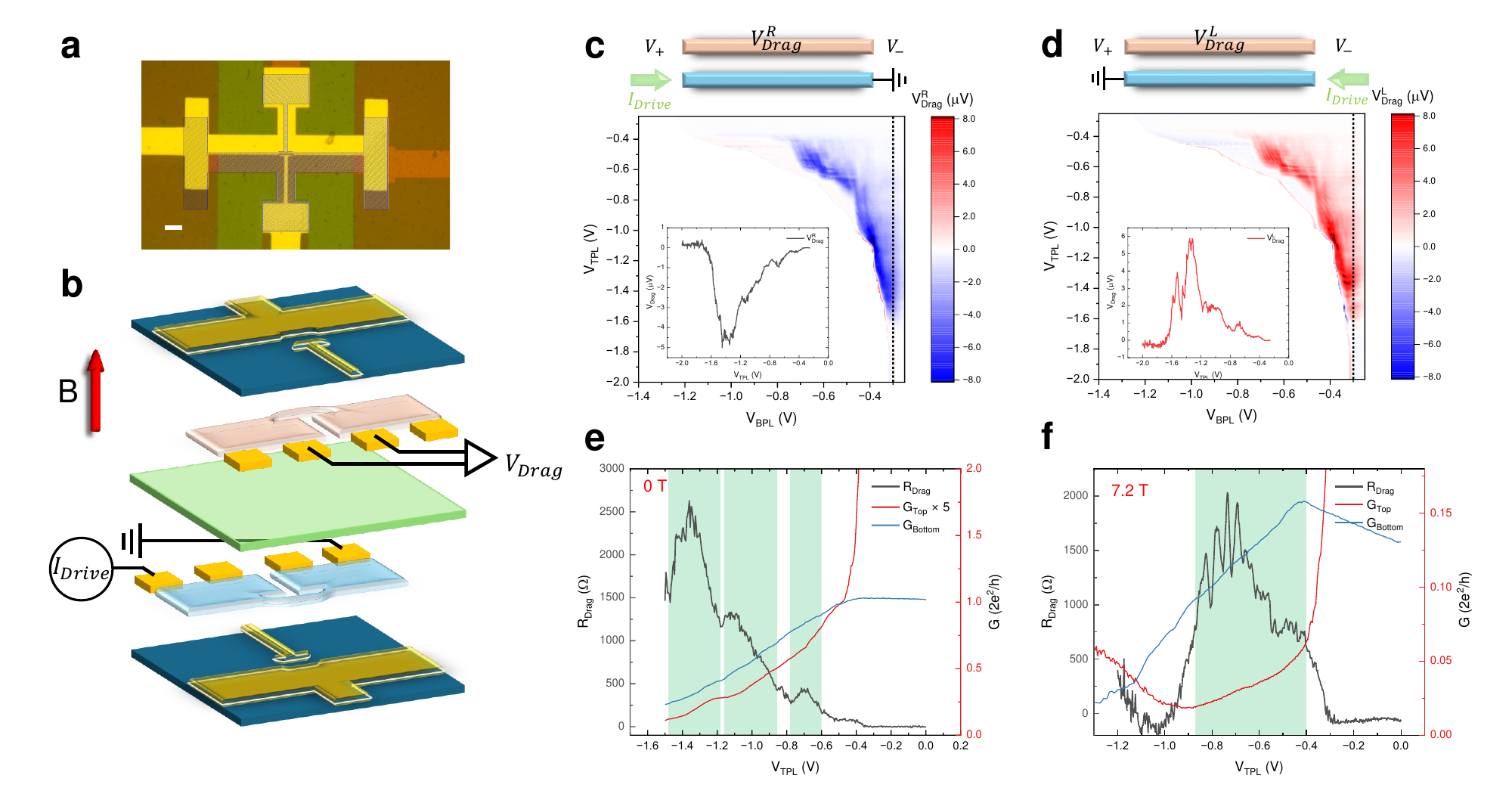}
\caption{\textbf{Onsager relation and Drag signal subband dependence.} \textbf{a}, Microscope image of the device. The scale bar is 10 $\mu m$ .\textbf{b}, Schematic of device structure with negative voltage applied to four gates (shown in gold color). Each wire is defined by one plunger (PL) and one pinch-off (PO) gate on each 2DEG. The PO gates are primarily used to independently contact the quantum wires and minimize tunneling current between them, while the PL gates are used to adjust the wires’ width and electronic density. The top plunger (TPL) and top pinch-off (TPO) gates form the top wire electrostatically on the top 2DEG, which is shown in pink. The yellow squares are the Ohmic contacts connecting the 2DEG. Similarly, the bottom wire defined on the bottom 2DEG is shown in blue. The two wires overlap each other and are separated by an AlGaAs barrier 15 nm wide (shown in green). In the following experiment, the top wire is used as the drag wire and the bottom wire as the drive wire, unless specifically noted. \textbf{c}, Drag voltage as a function of the top (drag) and bottom (drive) gate voltages when the drive current is in the same direction as the drag voltage measurement. The measurement was performed at the cryostat base electron temperature, below 15 mK. \textbf{d}, Same as \textbf{a}, but with the drive current direction reversed. The insets of \textbf{c} and \textbf{d} show the drag voltage at line 1 at $V_\text{BPL} = -0.3 \text{ V}$ in \textbf{c} and \textbf{d}, respectively, as shown by the dotted lines. \textbf{e}, Drag resistance as a function of TPL gate voltage at line 1, compared with the drag and drive wire conductance at 0 T. Three green shaded boxes represent the positions of the $1^\text{st}$, $2^\text{nd}$, and $3^\text{rd}$ subbands of the drag wire from left to right. \textbf{f}, Same as \textbf{e}, at 7.2 T. The upturn of the drag resistance and top wire conductance for $V_\text{TPL} < -0.9 \text{ V}$ corresponds to the tunneling signal when the wire is over-depleted and is not discussed further.  For both 0 T and 7.2 T, the main drag signal peak is located at the depletion position of the drag wire $1^\text{st}$ subband.}\label{fig1}
\end{figure}
\endgroup

\begingroup
\begin{figure}[h!]
\centering
\includegraphics[width=1\textwidth]{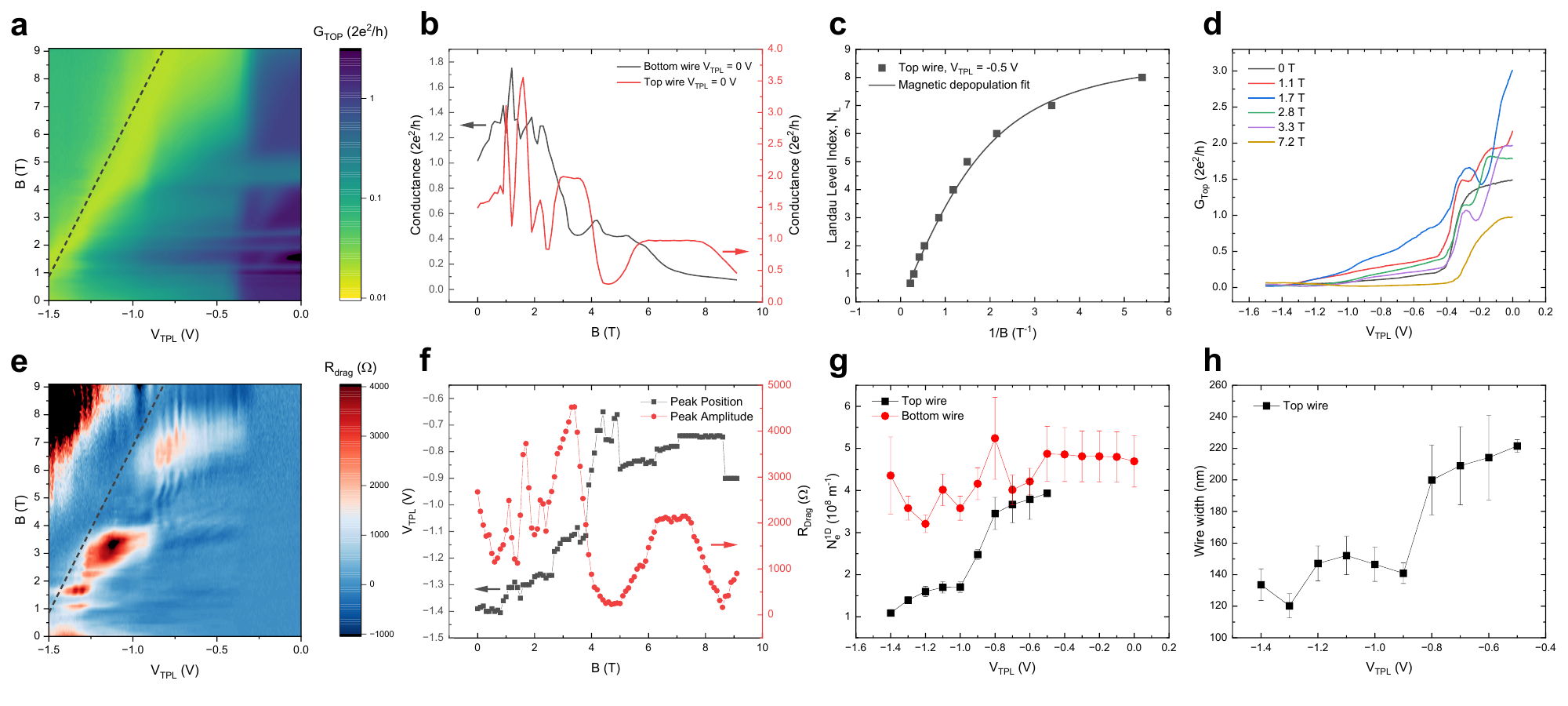}
\caption{\textbf{Coulomb drag between magnetic depopulated wires.} \textbf{a}, 2D map of the top (drag) wire conductance as a function of TPL gate voltage and magnetic field. The conductance signal oscillates as a function of the magnetic field, representing the magnetic depopulation of different wire 1D subbands. The dark gray dashed line in \textbf{a} and \textbf{e} represents the shift of the drag wire’s pinch-off position due to magnetic depopulation. The signal on the left of the pinch-off line is attributed to tunneling and not discussed here. \textbf{b}, Oscillatory magnetoconductance of the top and bottom wires at $V_\text{TPL} = 0 \text{ V}$, $V_\text{BPL} = -0.3 \text{ V}$ on line 1. The residual oscillatory minima observed in the top wire at $V_{\mathrm{TPL}}=0~\mathrm{V}$ originate from weak parabolic confinement induced by the over-pinched-off BPO gate (see Supplementary Note 3). \textbf{c}, Plot of Landau level index versus $B^{-1}$ for the magnetoconductance traces from \textbf{d}. Deviations from a straight line indicate subband depopulation effects. The solid line represents a fit of the parabolic confinement model. \textbf{d}, Top wire conductance of a few magnetic fields near the oscillation peaks in \textbf{f} on line 1. \textbf{e}, 2D map of drag resistance as a function of TPL gate voltage and magnetic field. The $1^\text{st}$ subband drag resistance peak oscillates and shifts towards more positive gate voltages as the magnetic field increases. \textbf{f}, Tracked 1st subband drag peak amplitude and position as a function of magnetic field. The drag signal oscillation positions in \textbf{f} overlap exactly with the drag wire magnetic depopulation position, as shown in \textbf{b}.  \textbf{g}, Fitted top and bottom wires 1D density at different TPL gate voltages in line 1. Both the top wire width and density are reduced with $V_\text{TPL}$, while the bottom wire density also decreases due to interwire crosstalk. \textbf{h}, Fitted top wire width at different TPL gate voltages in line 1. The bottom wire width varied minimally with TPL. The errors in \textbf{g} and \textbf{h} were calculated using a bootstrap Monte Carlo method using the half of the magnetic field step size as the error standard deviation.}\label{fig2}
\end{figure}
\endgroup

\begin{figure}[h!]
\centering
\includegraphics[width=0.3\textwidth]{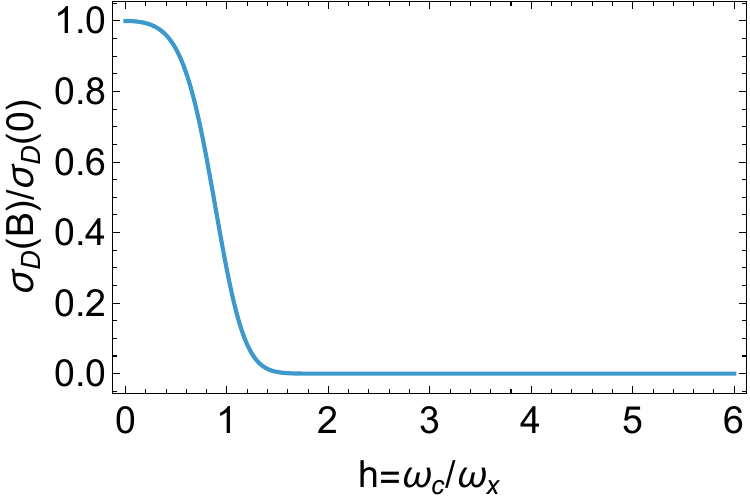}
\includegraphics[width=0.3\textwidth]{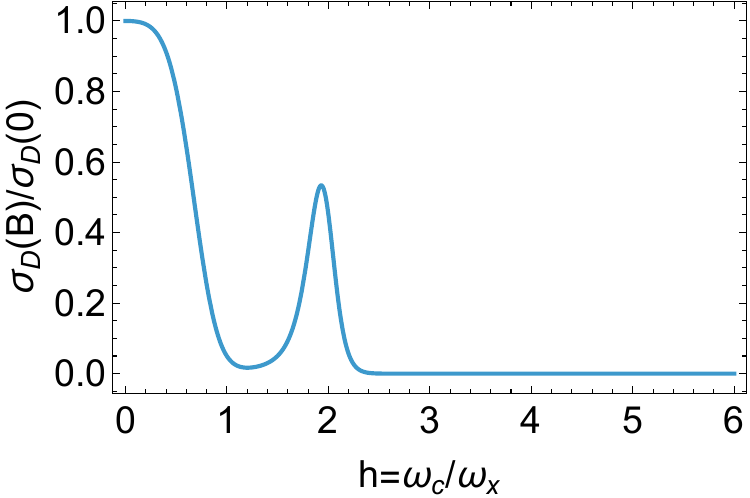}
\includegraphics[width=0.3\textwidth]{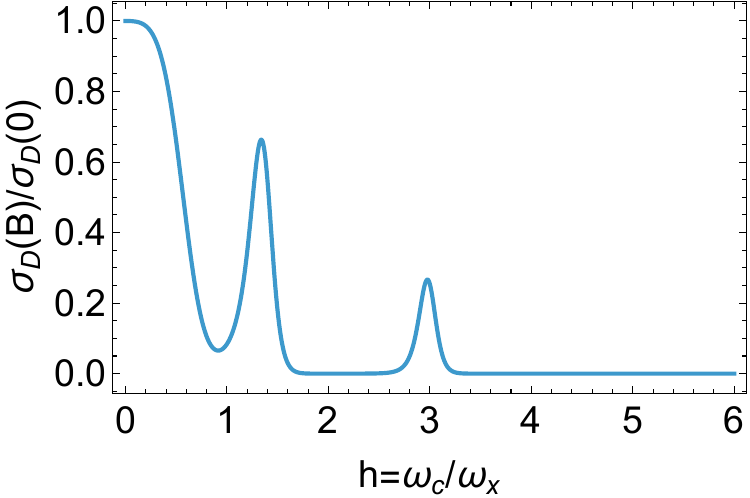}
\includegraphics[width=0.3\textwidth]{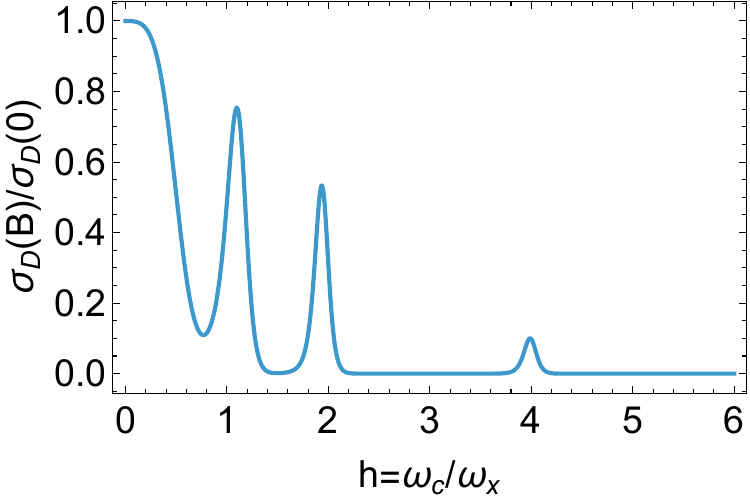}
\includegraphics[width=0.3\textwidth]{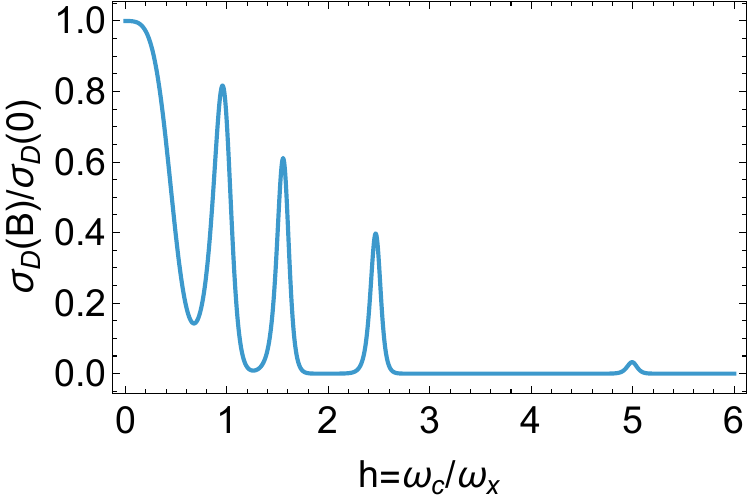}
\includegraphics[width=0.3\textwidth]{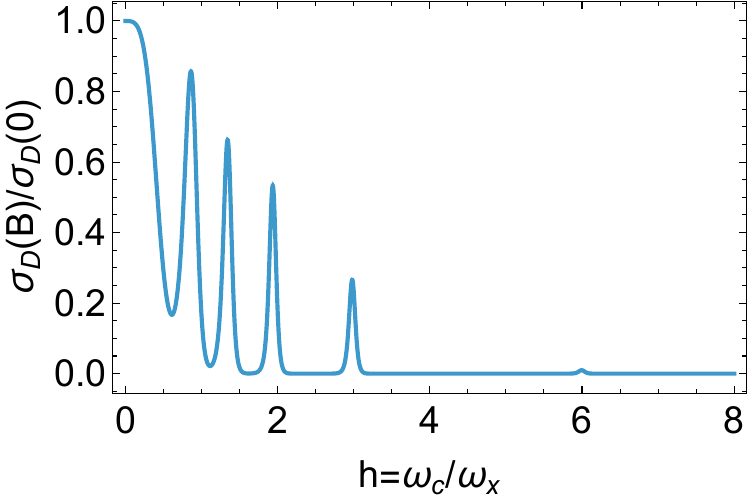}
\caption{\textbf{Model results for Coulomb drag magneto-oscillations.}
Normalized drag conductance as a function of magnetic field, $\sigma_D(B)/\sigma_D(0)$, computed from the transmission eigenvalue model of Ref. \cite{Fertig_1987,Buttiker_1990}. Mode-dependent transmission coefficients were computed based on the globally adiabatic saddle-point model of a wire constriction. Each panel corresponds to the same fixed value of a ratio $\omega_y/\omega_x=1$ between frequencies corresponding to the confining curvature and saddle curvature of the potential respectively. The dimensionless scale of the field $h$ is described by the cyclotron frequency $\omega_c$ measured in units of $\omega_x$. Gate voltages in each wire $eV_{g1,2}$ are also measured in units of $\omega_x$. In this model calculation, one gate voltage is kept fixed while the other is varied between $eV_{g1}/\omega_x = 1, 2, \dots, 6$, producing each of the six panels. The peaks appearing in the drag conductance oscillations correspond to field-induced sub-band transitions. Similar oscillations arise without a magnetic field by varying the gate voltage in each of the wires.
}\label{figGdrag}
\end{figure}

\begin{figure}[h]
\centering
\includegraphics[width=1\textwidth]{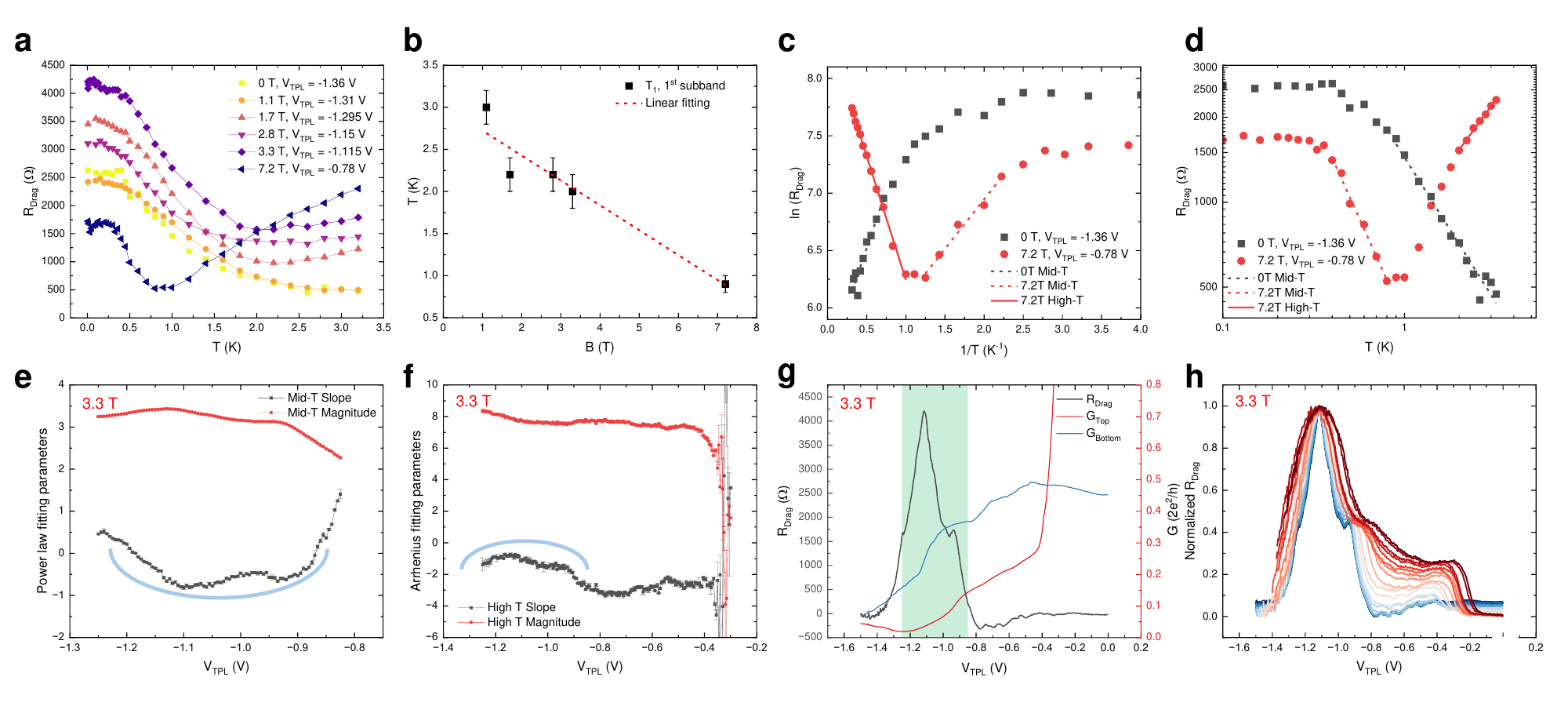}
\caption{\textbf{Temperature-dependent Coulomb drag at various magnetic fields.} \textbf{a}, $1^\text{st}$ subband peak drag resistance as a function of temperature at various magnetic fields in a linear scale. \textbf{b}, Upturn transition temperature $\text{T}_1$ as a function of magnetic field for the $1^\text{st}$ subband peak. With the magnetic field increasing, $\text{T}_1$ decreases from 3 K to 1 K, and the dashed line shows the linear fitting of $\text{T}_1$ as a function of B. \textbf{c, d}, $1^\text{st}$ subband peak drag resistance as a function of temperature at 0~T and 7.2~T in \textbf{c}, log-log scale, and \textbf{d}, Arrhenius scale. The dashed lines in \textbf{c} and \textbf{d} are the linear fitting of the intermediate temperature (Mid-T) regime, while the solid lines are the linear fitting of the high temperature (High-T) regime. For the Mid-T regime, drag resistance can be fitted better with a larger range with the power law. For the High-T regime, drag resistance can be fitted better with a larger range with the Arrhenius model. \textbf{e}, Power-law fitting result of temperature-dependent drag resistance in the Mid-T regime at line 1 at 3.3 T. The power law exponent has a minimum value of -1 at the drag peak position, while it increases on two sides of the drag peak, with the intercept of the opposite trend, reflecting the drag resistance peak shape. \textbf{f}, Arrhenius fitting result of temperature-dependent drag resistance in the High-T regime at line 1 at 3.3 T. Opposite to the power law exponent, the Arrhenius exponent has a maximum value of -1 at the drag peak position, while it decreases to -2 on both sides of the drag peak. The upturn of the Mid-T power law exponent and the downturn of the High-T Arrhenius exponent both reflect the broadening of the drag signal peak with temperature increasing, as shown in \textbf{h}. \textbf{g}, Drag resistance as a function of TPL gate voltage at line 1, compared with the drag and drive wire conductance at 3.3 T. The green shaded box represents the position of the $1^\text{st}$ subband of the drag wire. \textbf{h}, Normalized drag resistance peak as a function of temperature, at 3.3 T. The color, ranging from blue (light shades) to red (dark shades), corresponds to the temperature from 7 mK to 3.2 K. All the line traces are shifted to the one at base temperature to compare the drag peak width. 
}\label{fig3}
\end{figure}

\begin{figure}[h!]
\centering
\includegraphics[width=1\textwidth]{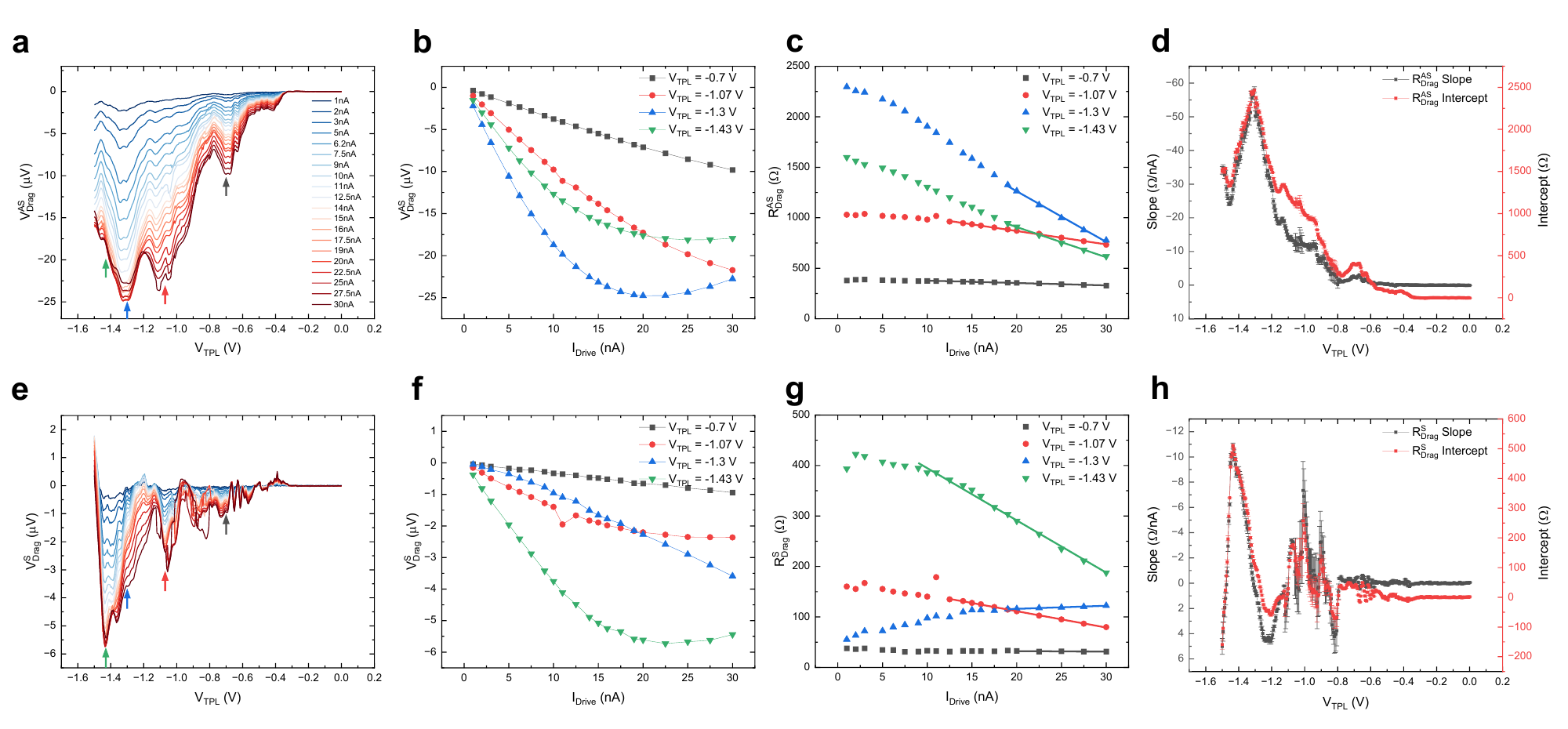}
\caption{\textbf{Nonlinear I-V and linear I-R at 0 T.} \textbf{a}, Antisymmetric drag voltage $V^\text{AS}_\text{Drag}$ as a function of $V_\text{TPL}$ and drive current $I_\text{Drive}$ for line 1. $I_\text{Drive}$ ranges from 1 nA to 30 nA. \textbf{b}, Nonlinear and nonmonotonous $V^\text{AS}_\text{Drag}$ as a function of $I_\text{Drive}$ at a few selected peak positions, shown by the arrows in \textbf{a}. \textbf{c}, Linear $R^\text{AS}_\text{Drag}$ as a function of $I_\text{Drive}$ at the selected peak positions. The solid lines represent the linear fittings in the high-bias regime. \textbf{d}, Linearly fitted slope and intercept of $R^\text{AS}_\text{Drag}$ as a function of $V_\text{TPL}$. \textbf{e-h}, similar to \textbf{a-d}, but for the symmetric component of drag signals. While both $\text{V}^\text{AS}_\text{Drag}$ and $V^\text{S}_\text{Drag}$ have strong nonlinearity and even nonmonotonic dependencies, $R^\text{AS}_\text{Drag}$ and $R^\text{S}_\text{Drag}$ evolve linearly in the high bias regime. Notably, the fitted slope has the exact same gate dependence as the $R_\text{Drag}$ at a low bias of 2 nA. 
}\label{fig4}
\end{figure}







\end{document}